\documentclass{svmult}

\usepackage{mathptmx}       
\usepackage{mathtools}
\usepackage{amssymb}
\usepackage{slashed}
\usepackage{helvet}         
\usepackage{courier}        
\usepackage{type1cm}        
\usepackage{makeidx}         
\usepackage{graphicx}        
\usepackage{multicol}        
\usepackage[bottom]{footmisc}
\usepackage{hyperref}        
\usepackage{soul}            
\hypersetup{colorlinks=true,urlcolor=blue}
\usepackage[numbers,sort,compress]{natbib}
\makeindex             

\usepackage{xcolor} 
\definecolor{darkgreen}{rgb}{0.01, 0.75, 0.24} 
\usepackage{tensor}
\newcommand{\tnsr}[2]{\tensor{#1}{#2}} 
\newcommand{\eref}[1]{(\ref{#1})}
\def\be#1\ee{\begin{align}#1\end{align}}
\def\bsub#1\esub{\begin{subequations}#1\end{subequations}}
\def\nn{\nonumber}
\def\q{\qquad}
\def\f{\frac}
\def\eps{\varepsilon}

\def\lb{\big\lbrace}
\def\rb{\big\rbrace}
\def\time{\,{\scriptstyle{\times}}\,}

\def\de_\omega{\mathrm{D}}
\def\de{\mathrm{d}}

\def\E{\mathcal{E}}

\def\I{\mathcal{I}}

\begin{document}
\title*{Corner symmetry and quantum geometry}

\author{Laurent Freidel, Marc Geiller, Wolfgang Wieland}

\institute{Laurent Freidel \at Perimeter Institute for Theoretical Physics\\ 31 Caroline Street North, Waterloo, Ontario, Canada N2L 2Y5\\ \email{lfreidel@perimeterinstitute.ca}
\and Marc Geiller \at ENS de Lyon, CNRS, Laboratoire de Physique, F-69342 Lyon, France\\ \email{marc.geiller@ens-lyon.fr}
\and Wolfgang Wieland \at Institute for Quantum Gravity, Theoretical Physics III\\
Friedrich-Alexander Universität Erlangen-Nürnberg\\
Staudtstrasse 7/B2, 
91058 Erlangen,
Germany\\
Institute for Quantum Optics and Quantum Information (IQOQI)\\ Austrian Academy of Sciences, Boltzmanngasse 3, 1090 Vienna
Austria \at Vienna Center for Quantum Science and Technology (VCQ)\\ Faculty of Physics, University of Vienna, Boltzmanngasse 5, 1090 Vienna, Austria\\ 
\email{wolfgang.wieland@fau.de}\\
\and\textit{This is a preprint of a chapter to appear in the “Handbook of Quantum Gravity”, edited by Cosimo Bambi, Leonardo Modesto and Ilya Shapiro, 2023, Springer, reproduced with permission of Springer.}}

\maketitle

\vspace{-3.5cm}

\abstract{
By virtue of the Noether theorems, the vast gauge redundancy of general relativity provides us with a rich algebra of boundary charges that generate physical symmetries. These charges are located at codimension-2 entangling surfaces called corners. The presence of non-trivial corner symmetries associated with any entangling cut provides stringent constraints on the theory's mathematical structure and a guide through quantization. This report reviews new and recent results for non-perturbative quantum gravity, which are natural consequences of this structure.
First, we establish that the corner symmetry derived from the gauge principle encodes quantum entanglement across internal boundaries. We also explain how the quantum representation of the corner symmetry algebra provides us with a notion of quantum geometry. We then focus our discussion on the first-order formulation of gravity and show how many results obtained in the continuum connect naturally with previous results in loop quantum gravity. In particular, we show that it is possible to get, purely from quantization and without discretization, an area operator with discrete spectrum, which is covariant under local Lorentz symmetry. We emphasize that while loop gravity correctly captures some of the gravitational quantum numbers, it does not capture all of them, which points towards important directions for future developments. Finally, we discuss the understanding of the gravitational dynamics along null surfaces as a conservation of symmetry charges associated with a Carrollian fluid.
}

\section*{Keywords} 
Boundary degrees of freedom, holography, Loop Quantum Gravity, Noether charges, symmetry algebras, corner symmetries, Carrollian geometry, representation theory of infinite-dimensional groups.

\section{Introduction}

Quantum gravity is one of our time's most captivating and challenging theoretical puzzles. Many different top-down approaches have been proposed to address it, such as\footnote{This is by no means a complete list. See for example \cite{deBoer:2022zka}.} String Theory, AdS/CFT Holography\footnote{Celestial holography can also be understood as a bottom-up perspective on quantum gravity that is perturbative in nature.}, and Loop Quantum Gravity (LQG hereafter). Despite many successes, these approaches are based on fundamentally different foundational principles and seem irreconcilable. What is missing, in our view, is a  bottom-up non-perturbative perspective that provides, from our understanding of semi-classical gravity and its symmetries, solid foundations and tests that any top-down approach to quantum gravity ought to pass. It includes one of the foundational elements of holography, namely the fact that the Hamiltonian of a gravitational system with fixed boundary conditions is entirely contained on its boundary. It also respects one of the fundamental tenants of LQG, namely background independence, and the foundations of the equivalence principle, which implies that gravitational dynamics is the expression of its symmetry. The local holography approach views both holography and symmetry as two sides of the same coin. In this review, we focus on presenting the connection of local holography with LQG and some of its results such as the appearance of quantum geometry, leaving aside a more general presentation of its link with traditional holography.

In particular, we show that gauge symmetry, which is usually understood as a mere redundancy of our physical description, is much more than that. It governs the quantum entanglement of subregions across entangling surfaces, called corners. These entangling surfaces are shown to carry representations of infinite-dimensional symmetry groups that are the physical consequence of the underlying gauge invariance. Moreover, we show that these symmetries represent, through their non-trivial commutation relations, the elements of quantum geometry.

 This perspective is very close to some of LQG's most noticeable technical achievements, in particular, the rigorous construction of a Hilbert space of quantum Riemannian geometry based on a Hamiltonian formulation of general relativity in terms of $\mathrm{SU}(2)$ Yang--Mills gauge connection variables. In this context, symmetries such as spatial diffeomorphisms and local $\mathrm{SU}(2)$ frame rotations were central for defining the physical states. The emphasis on SU$(2)$ gauge symmetry led to the quantum numbers labeling the kinematical spin network states \cite{Haggard:2023tnj}. 

Since it was initially conceived as a canonical quantization of general relativity in the Hamiltonian formulation, LQG is often seen as a bulk quantization. Little to no role is played by boundary structures. One goal of the present review chapter is to explain why this is not so and how LQG accommodates the description of some boundary degrees of freedom in its foundations. For example, one of the key results of LQG is that spin networks represent a notion of quantum geometry. Suppose we cut open a portion of quantum geometry along a surface $S$. In that case, we obtain, for each intersection between the spin network and the surface $S$, an $\mathrm{SU}(2)$ representation, which carries an elementary quantum of geometry. Moreover, the $\mathrm{SU}(2)$ generators associated with each cut are the quantum flux operators integrated along $S$. These quanta, which are fundamentally entangled and represent quantum geometry, are controlled by gauge symmetry. One limitation of the traditional LQG approach is that it assumes a discretization of the fundamental geometrical excitations. Local holography allows us to keep some of the key successes of LQG, such as the quantization of the area spectra while being fully compatible with the standard understanding of QFT in the continuum. The presentation will be done in four main parts.

In section \ref{sec:gauge}, we first explain how the notion of boundary degrees of freedom arises in gauge theories, and highlight the fundamental role that it plays in classical and quantum gravity. We develop these ideas further in section \ref{sec:quantum geo} by explaining how boundary symmetries are related to entanglement and how their quantization gives rise to a notion of quantum geometry. 
This then motivates us to summarize the role played by boundary degrees of freedom in LQG.
We recall in section \ref{sec:old} various results establishing that spin network data can be seen as boundary data and also summarize earlier results on boundaries and entanglement in LQG. We then review recent results on the construction of corner charges in the first-order formulation of gravity in the presence of the Barbero--Immirzi parameter. And give the proof of a Lorentz covariant area operator with discrete spectra. This section also reveals our shared viewpoint on future developments. It shows how the non-perturbative quantum gravity perspective started by LQG should be extended in order to fully provide a representation of the boundary symmetries compatible with subsystem decomposition and coarse-graining.
Finally, in section \ref{sec:null}, we review our current understanding of the representation of supertranslation symmetry, emphasizing the gravitational dynamics along null boundaries. We present our understanding of such dynamics as Carrollian charge conservation equations. We also construct a boundary action for finite null boundaries that encodes radiation and is compatible with discrete area spectra.

\section{Gauge theories in bounded regions}
\label{sec:gauge}

\subsection{Generalized Noether charges and boundary symmetries}

In the case of gauge symmetries, Noether's first and second theorems together imply that charges are associated to codimension-2 boundaries. More precisely, to local gauge symmetries with parameter $\eps$ one can assign the object
\be\label{generalized Noether charge}
Q(\eps)=\int_\Sigma C(\eps)+\oint_Sq(\eps),
\ee
where $C(\eps)\hat{\;=\;}0$ is the constraint and $q(\eps)$ the so-called \emph{charge aspect}. This result implies that the charges of local symmetries supported inside a region $\Sigma$ are given by an integral over its boundary $S=\partial\Sigma$. By definition, what then distinguishes a gauge transformation from a true physical symmetry is the possible non-vanishing value of the charge. We therefore see that in the presence of a boundary, whose intersection with $\Sigma$ is the corner $S$, gauge transformations can potentially become physical symmetries associated with a non-vanishing charge. Setting aside technical issue for the moment, one can think of the covariant phase space formalism as a method for assigning the charge \eqref{generalized Noether charge} to an infinitesimal transformation $\delta_\eps$.

Such surface charges are of fundamental importance as they describe essential physical observables. This is the case of the ADM mass, the angular momentum, the memory observables, the multipole moments, and for example the LQG fluxes. Moreover, the charges come equipped with certain algebraic structures: they form a local symmetry algebra, and their time development along the boundary can give rise to so-called flux/balance laws, describing for example the interplay between gravitational radiation and the failure of the charges to be conserved. It is indeed important to appreciate that the physically meaningful surface charges go beyond what one can be used to when studying Noether charges in mechanics. The charges can be non-conserved, and also defined without actually resorting to boundary conditions, i.e.\ by relaxing the need for a variational principle. Moreover, the charges can also fail to be integrable in the variational sense.

For example, in the case of a diffeomorphism with parameter $\xi$, the charge takes the form
\be\label{general diffeo charge}
I_\xi\Omega=\slashed{\delta}Q(\xi)=\delta Q_\text{N}(\xi)-\oint_Si_\xi\theta,
\ee
where $\Omega$ is the symplectic structure, obtained as the integral over $\Sigma$ of the variation $\delta\theta$ of the symplectic potential $\theta$. This latter appears as the boundary term of the variation of the Lagrangian, i.e.\ $\delta L\hat{\;=\;}\de\theta$. Finally, here $Q_\text{N}(\xi)$ stands for the Noether charge associated with the Lie derivative $\pounds_\xi$. Importantly, this formula shows that the full surface charge is not only given by the Noether piece, but can contain a non-integrable flux contribution $F_N(\xi) = \oint_S i_\xi \theta$. More generally, the  charge variations take the form
\be
I_\eps\Omega=\delta Q(\eps)+F(\eps),
\ee
where the right-hand side is a (non-unique) split between an integrable piece $Q(\eps)$ and a flux contribution $F(\eps)$. We leave aside technical discussions on how to select a preferred split of the charge between integrable and non-integrable parts, but a canonical choice is the Noetherian split, $Q(\eps)=Q_\text{N}(\eps)$, which is entirely determined by the choice of boundary Lagrangian \cite{Freidel:2020xyx, Freidel:2021cjp}. What is important is that a repeated symmetry action on a charge leads to the equation
\be\label{Charge-Flux}
\delta_{\eps_2}Q(\eps_1)&=Q([\eps_1,\eps_2])+I_{\eps_1}F(\eps_2)\cr
&\ \ \rotatebox[origin=c]{90}{$\Leftrightarrow$}\\
\text{evolution}&=\text{conservation}+\text{dissipation}.\nn
\ee
As indicated, this can be thought of as an evolution generated by the symmetry, which is composed of a rotation dictated by the symmetry algebra, and a possible dissipation due to the flux term. This latter is typical of open Hamiltonian systems, in which modes can escape e.g. through the null boundary in the form of gravitational radiation. An important point of this construction explained in details for the first time  in \cite{Donnelly:2016auv} is that once we take into account the presence of edge mode fields, it is possible to show that the symmetry charges are gauge invariant. They commute with the Hamiltonian constraints. In Gravity, the edge mode fields are given by the choice of embedding maps from a reference ball to the spacetime \cite{Freidel:2021dxw, Ciambelli:2021nmv}. For simplicity, we do not introduce them in our presentation even if they play an important conceptual role. 

This structure suggests that there is both kinematical and dynamical information in the charges and their algebra. Kinematical charges are integrable and associated with a vanishing flux, while the dynamics amounts to restoring the flux and taking into account the radiative modes. The kinematical charges belong to the so-called \emph{corner symmetry group}, which is a subgroup $G_0^S\subset G^S_\text{ext}$ of the full symmetry group.  The corner symmetry charges form a closed algebra $\{Q_0(\eps_1),Q_0(\eps_2)\}=Q_0([\eps_1,\eps_2])$ which can be readily quantized. The full symmetry group is called  the \emph{extended corner symmetry group}. The extended corner symmetry charges can only be understood as canonical symmetry charges after extending the gravitational phase space with an edge mode field called the embedding field \cite{Freidel:2021dxw, Ciambelli:2021nmv}. More recent work \cite{Kabel:2022efn} connects this research to the geometry of open systems, dissipation and entropy production via the framework of metriplectic geometry. At the quantum level, physical observables should provide representations of $G_0^S$ and $G^S_{\mathrm{ext}}$.   This is a necessary condition in order to have a quantization compatible with the symmetries of the underlying classical theory. In general relativity, as we will show below, representing the corner charge algebra amounts to a quantization of geometry, while representing the full dynamical charge algebra is expected to provide a notion of quantum dynamics. For a demonstration of this in the context of asymptotic infinity see \cite{Freidel:2021qpz}.

What we have left out of the discussion so far is an explanation of the zoology of charges and charge algebras which can arise. This depends of course on $i)$ the theory being considered (e.g. gravity or Maxwell), $ii)$ the symmetry transformations being studied (e.g. diffeomorphisms or internal Lorentz transformations),  $iii)$ the type of boundary conditions (e.g. conservative or radiative boundary conditions), $iv)$ the nature of the boundary (e.g. an asymptotic  boundary or  an entangling surface at finite distance), $iv)$  The choice of boundary Lagrangian or topological terms such as theta terms.
Recent years have seen a wealth of developments and new results on  the covariant phase 
 space which has fueled lots of the developments we present in this survey. In addition to the references \cite{Donnelly:2016auv,Freidel:2020xyx, Freidel:2021cjp,Freidel:2021dxw, Ciambelli:2021nmv} cited above the interested reader should consult  
 \cite{Oliveri:2019gvm,Chandrasekaran:2020wwn,Chandrasekaran:2021vyu,Margalef-Bentabol:2020teu,Odak:2022ndm,Wieland:2021eth}.  Moreover, the recent report view \cite{Ciambelli:2022vot} contains a comprehensive review of the ideas we presented here and many more references.

In order to make contact with LQG, we will focus below on gravity in tetrad variables, and consider the corner symmetries and charges which can be derived for any quasi-local region (i.e.\ a boundary at finite distance) and without the need for boundary conditions. An immediate example of such a charge is that obtained from \eqref{general diffeo charge} in the case of tangential vector fields, for which the flux term vanishes.\footnote{One should distinguish between symplectic flux, which can exist at a cut, from the flux due to time evolution. In the case of tangential vector fields the former is absent, but of course there is still a non-trivial flux of e.g. angular momentum due to its time evolution along $\I$. There is therefore a translation flux but no rotation flux. On equation \eqref{Charge-Flux}, one can indeed see that even if $\eps_1$ is tangent, then there is still a non-trivial flux due to time evolution if $\eps_2$ is an infinitesimal translation in the $u$ direction.}

\subsection{Implications for classical and quantum gravity}

The role of symmetries in the study of a classical and quantum theory cannot be understated. Provided enough conserved quantities exist in a theory, they can be used to solve the dynamics altogether. Although we do not expect this to be the case in gravity, it is still the case that symmetries provide a powerful organizational tool, and give the possibility of controlling the quantization of geometry in terms of symmetry observables.


In the case of asymptotic symmetries for asymptotically-flat spacetimes, the relationship between corner symmetries (described by the so-called BMS group and its generalizations \cite{Bondi,Sachs}), classical observables, and the structure of the perturbative quantum theory, has already been demonstrated, and is being used to build a notion of flat space holography (see \cite{Strominger:2017zoo,Raclariu:2021zjz,Pasterski:2021rjz} and references therein). This idea of building the quantum theory on the structure of the classical symmetries is also what gave rise to the holographic AdS/CFT correspondence. 
It is clear that the very existence of boundary charges and symmetries in the classical theory should play a central role in any attempt to quantize gravity and understand what form of holography holds in quantum gravity.

\section{Quantum geometry and corner symmetry}
\label{sec:quantum geo}

The viewpoint just presented  helps understand the fundamental tension which exists between gravity and local quantum field theory. In any gauge theory, the presence of constraints implies that gauge-invariant observables are necessarily non-local. This means, in turn, that the Hilbert space does not factorize spatially, as factorization would leave out of the sub-factors some of the non-local observables which cut across the entangling surface (see \cite{Buividovich:2008gq,Casini:2014aia,Donnelly:2011hn,Donnelly:2014gva,Donnelly:2015hxa,Geiller:2019bti,Gomes:2018dxs} for many discussions on this issue in gauge theory).

\subsection{Entanglement from symmetry}

Let us now explain more in details, why corner symmetries are essential to unravel the nature of spacetime entanglement and lead to a covariant description of quantum spacetime that generalises the one given by LQG.
The key idea proposed in \cite{Donnelly:2016auv} is that corner symmetry allows us to resolve the quantum entanglement of spacetime through symmetries derived from the gauge principle. This correspondence between symmetry and entanglement was first realised in the context of lattice gauge theory in  \cite{Donnelly:2014gva}. 

Let us start with an example in standard field theory and let us focus on the vacuum state $|0\rangle$. This state is an eigen state of the Hamiltonian $H$ but also of the boost operator $K$ that fixes the 2-dimensional plane $S=\{x^3=0\}$   inside the $x^0=\mathrm{const}.$ Cauchy slice. In other words we have 
\be 
H|0\rangle= K|0\rangle =0.
\ee 
The plane $S$ fixed by $K$ is called the \emph{entangling surface}. It divides the slice into a left and a right component $\Sigma=\Sigma_L\cup \Sigma_R$ with $S=\partial \Sigma_{L/R}$. 
One can decompose the Hilbert space as a product of left (resp.  right) Hilbert spaces\footnote{Strictly speaking this decomposition is only valid at the level of the observable algebras due to UV divergences. See \cite{Witten:2018zxz} for the proper description. For simplicity, we keep the naive language of Hilbert space factorisation, valid for a UV-regulated QFT. } obtained by acting on the vacuum by local operators supported on $\Sigma_L$ (resp. $\Sigma_R$).
Tracing out the left Hilbert space one obtain the density matrix $\rho_S=\mathrm{Tr}_{H_L}(|0\rangle\langle0|)$.

Since the entangling cut is fixed by $K$, and the slice is defined at $t=0$, we can decompose the total boost  operator into a local sum $K=K_R-K_L$, where $K_{R/L}=\int_{\Sigma_{R/L}} |x^3| T_{00}$ is the right/left boost operator, also called the ``twist operator'',  which commutes with local  operators placed on the left side.
$K_{R/L}$ is a positive operator, and it  acts non trivially on the vacua.This means that we can decompose the density matrix in terms of eigenstates $|n\rangle $ of the ``twist operator'' $K_R$ with eigenvalue $K_n>0$. 

We can now use these symmetry generators to construct the \emph{symmetry-reduced} state which are defined as the projection of $\rho_S$ onto  the eigenspaces of $K_R$.
The main theorem of Unruh--Bisognano--Wichmann \cite{Bisognano:1976za} states that the symmetry-reduced vacuum is pure! The second statement of this theorem is that this state is simply given by a complex boost associated with an imaginary boost angle of $2i\pi$. Let us recall that U$(\tau)= e^{\frac{i}{\hslash} \tau K_R}$ denotes the action of the boost operator with boost angle $\tau$ on states supported on $\Sigma_R$. The Unruh--Bisognano--Wichmann result therefore simply states that 
\be 
\rho_S = e^{-\frac{2\pi}{\hslash} K_R}.
\ee
This means that the vacuum state can be decomposed as
\be
|0\rangle =\sum_n e^{-\frac{\pi}{ \hslash} K_n} |n, \tilde n\rangle,
\ee 
where $|n,\tilde{m}\rangle$ denotes an eigen state of $K_R$ (resp. $K_L$) with eigenvalue $K_n$ (resp. $K_m$).
The lesson we learn from this example is that the vacuum entanglement is entirely contained from the matching of boost charges $K_R=K_L$ across the entangling surface. This is the expression of boost invariance which implies that the state can be written as $\sum_n \rho_n |n,\tilde{n}\rangle$ in term of a measure $\rho_n$.  We also learn that the form for this measure  is determined by the demand of translation invariance of the state.

So far we have reviewed a well-known result about quantum field theory (QFT) in the presence of an entangling surface. How does this relate to quantum gravity and corner symmetries? 
In QFT we saw that the only state to which we can apply the previous symmetry argument to compute entanglement is the vacuum state for which 
$K|0\rangle=0=H|0\rangle$. In gravity we know, as reviewed in the previous section, that on a closed Cauchy surface the Hamiltonian generator $Q(\xi)$ vanishes for \emph{any} vector field $\xi$.\footnote{A weaker statement would be that $\Sigma$ has an asymptotic boundary but $\xi$ vanishes sufficiently fast near infinity.} At the quantum level this means that $Q(\xi) |\Psi\rangle=0$ for \emph{all} quantum states of quantum gravity. Let us now choose a 2-dimensional finite closed surface $S\in \Sigma$ which plays the role of an entangling surface and denote $\Sigma_L$ the surface inside $S$ and $\Sigma_R$ the slice outside $S$. Let us then consider the set of vector fields $\xi$ which do not move $S$. 
We can then decompose the diffeomorphism charge in terms of its left and right components $Q(\xi)= Q_R(\xi)-Q_L(\xi) $. 
As we have seen in the previous section, $Q(\xi)$ is the integral of the Hamiltonian constraint on $\Sigma$ while  the charges $Q_R(\xi)$ are given by corner integrals and the corner symmetry algebra $\mathfrak{g}^S$ is the sub-algebra of infinitesimal diffeomorphisms fixing $S$ that possess a non vanishing charge corner charge $Q_R(\xi)=\oint_S q_\xi$. We are thus in a situation which is analogous to the Unruh case. We expect that the physical Hilbert space splits into a sum of representations $\rho$ of the corner symmetry group. In other words, given the existence of a corner symmetry algebra, the quantum gravity states $\Psi$ can be decomposed into a sum
\be
\Psi = \sum_\rho \psi_\rho(S) \sum_a |\rho , a\rangle \otimes  |\rho^* , a\rangle.
\ee
Here $\rho*$ denotes the dual representation and $a$ denotes a label of states in the Hilbert space $V_\rho$.
This decomposition can be extended to more than one surfaces $S\to (S_1,\cdots S_n)$ and more generally to a collections of cuts 
$\Gamma$ that represents a 2-dimensional cellular complex. The previous calculation of charges as corner charges generalises to this case. On the 2-dimensional cell of this complex we assign representations of the corner symmetry group and on the 1-dimensional cells which glue the 2-cells we assign intertwiners of the corner symmetry algebra. At the end of the day, we can therefore expect a decomposition of the quantum gravity state as a sum
\be 
|\Psi\rangle =
\sum_{\rho_\Gamma} \psi_{\rho_\Gamma}(\Gamma) \sum_{a_\Gamma} |\rho_\Gamma , a_\Gamma\rangle \otimes  |\rho_\Gamma^* , a_\Gamma\rangle.
\ee  
Here $\rho_\Gamma=(\otimes_f \rho_f,\otimes_e I_e)$, where $f$ labels the faces of $\Gamma$ with representations $\rho_f$, and $e$ labels the edges of the complex with intertwiners $I_e$. This decomposition must be valid for \emph{any} quantum gravity state.

This decomposition is reminiscent of the spin network decomposition of quantum gravity states that appears in loop quantum gravity\footnote{See \cite{Ashtekar:2004eh} or the chapter \cite{Haggard:2023tnj} in this volume.}, but it is also more general. The difference is that it appears in a continuum description and the group which one needs to represent is the full corner symmetry group (and not only the SU$(2)$ group of internal frame rotations as in LQG). As we are about to see, in the corner symmetry proposal,  the quantization of geometry appears through the choice of representation and through quantization of the spectra of the geometric operators that represents the corner symmetry charges. The central questions of quantum gravity now become, in this context: what are the representations which enter the decomposition of the quantum gravity states, and what is the integration measure $\Psi_\rho$ which plays the role of the Unruh measure $e^{\frac{\pi K_n}{\hslash}}$ for quantum gravitational states?

\subsection{Representation theory of corner symmetries}
\label{sec:representation}

In metric gravity it has been shown that the corner symmetry group\footnote{The extended corner symmetry group is obtained by adding the super-translations along the two normal directions, and is given by \cite{Donnelly:2016auv,Ciambelli:2021vnn}
\be 
G_{\mathrm{metric}, \mathrm{ext}}^S = \left(\mathrm{Diff}(S)\ltimes \mathrm{SL}(2,\mathbb{R})^S\right)\ltimes \mathbb{R}^{2S}.
\ee}
is simply the semi-direct product \cite{Donnelly:2016auv}
$ 
G_{\mathrm{metric}}^S = \mathrm{Diff}(S)\ltimes \mathrm{SL}(2,\mathbb{R})^S,
$
where $\mathrm{SL}(2,\mathbb{R})^S$ denotes the space of functions from the corner $S$ to SL$(2,\mathbb{R})$. These represent the boost transformations preserving $S$, while Diff$(S)$ are the sphere diffeomorphisms. In the first order tetrad (or frame) formalism, the symmetry group at the corner is extended by SL$(2,\mathbb{C})^S$, namely the group of internal local Lorentz transformations supported on the sphere \cite{Freidel:2020svx}. In the tetrad formulation the corner symmetry group is therefore
\be 
G_{\mathrm{frame}}^S =
\mathrm{Diff}(S)\ltimes\big(\mathrm{SL}(2,\mathbb{R})^S\times  \mathrm{SL}(2,\mathbb{C})^S\big).
\ee
The study of the representations of these groups has been initiated in the work \cite{Donnelly:2020xgu} and developed further in \cite{Freidel:2020svx,Freidel:2020ayo,Donnelly:2022kfs,Ciambelli:2022cfr}. 
To connect this general discussion to the traditional discussion in LQG, one can go back to the work \cite{Freidel:2019ees}, which proposed the study of the loop corner symmetry group 
\be 
G^S_\mathrm{LQG}= \mathrm{Diff}(S)\ltimes \mathrm{SU}(2)^S,
\ee 
obtained in the study of the canonical formulation of gravity in the time gauge. This is the subgroup of $G^S_{\mathrm{frame}}$ preserving the slice normal at $S$. In this context, the  gravitational phase space is described in the bulk in terms of the SU$(2)$ valued Ashtekar--Barbero connection $A^i$, and in terms of the 2-form flux element $E_i=\frac12\epsilon_{ijk} (e^j \wedge e^k)$, where  $e^i=\de x^a e_a^i$ denotes the frame field of the induced metric on $\Sigma$. Before imposing the constraints, these variables are canonically conjugated as
\be \lb A_a^i(x),E^b_j(y)\rb=\delta_a^b\delta^i_j\delta^{(3)}(x-y).\label{Comm}
\ee
The kinematical constraint equations associated with diffeomorphism and SU$(2)$ gauge symmetry are simply, as shown in \cite{Freidel:2019ees}, the Gauss conservation laws
\be 
\mathrm{d}_A E_i\hat{\;=\;}0,
\q\q
\mathrm{d}_A P_i\hat{\;=\;}0.
\ee 
Here $E_i$ is the geometrical 2-form flux, while $P^i\coloneqq \de_A e^i$ is the torsion of the Ashtekar--Barbero connection. On the reduced phase space $P^i$ computes components of the extrinsic curvature tensor. One can also understand $P_i$ as the variable conjugated to the frame field since
\be 
\lb P^a_i(x), e_b^j(y)\rb= \delta_b^a\delta^i_j\delta^{(3)}(x-y)
\ee on the kinematical phase space. 

It is important to appreciate that the commutation relations \eqref{Comm}, or the one just shown above, are only valid before imposition of the constraints. What matters at the quantum level is to understand how these variables commute in the \emph{reduced phase space} obtained after imposing the constraints.
This is where the corner symmetry enters as a key ingredient.
It is direct to show \cite{Freidel:2019ees, Freidel:2019ofr} that the symmetry charge of infinitesimal diffeomorphisms labelled by a vector $\xi^a\partial_a\in T\Sigma$ which is tangent to $S$ and of infinitesimal SU$(2)$ gauge transformation labelled by $\alpha^i\sigma_i \in \mathfrak{su}(2)$ are respectively given by  the corner charges
\be \label{LQG SU2 charges}
P(\xi) \hat{\;=\;} \oint_S\xi^ae^i_a P_i, \q\q G(\alpha) \hat{\;=\;}\oint_S E_i \alpha^i, 
\ee 
where the hatted equality emphasizes that these equation are valid on the reduced phase space.

In particular one sees that, after imposition of the Gauss constraints, the pull-back on $S$ of the electric field operator satisfies ultralocal commutation relations at the corner, is given by \cite{Cattaneo:2016zsq,Donnelly:2016auv}
\be \label{commE}
\lb E^i(\sigma), E^j(\sigma')\rb_S \;\hat{=}\;\epsilon^{ij}{}_k \delta^{(2)}(\sigma-\sigma') E^k(\sigma),
\ee 
where $\sigma,\sigma' \in S$. Usually such a non-trivial commutation relation of the flux operator is only obtained in LQG after the introduction of loops which discretize the support of the flux lines.  What is remarkable is that we can establish this non-commutation directly in the continuum and at the classical level through the Noether theorem.
Similarly, the momentum operators $P_A = P_i e^i_A$ with $A=1,2$ indices tangent to $S$, are the generators of diffeomorphisms along $S$. The commutation relations of these momenta on the reduced phase space are
\be \label{commP}
\lb P_A(\sigma),P_B(\sigma')\rb\;\hat{=}\;P_A(\sigma') \partial_B\delta^{(2)}(\sigma,\sigma')-P_B(\sigma) \partial'_A\delta^{(2)}(\sigma,\sigma'). 
\ee 

At the quantum level the goal is to represent these commutation relations. The commutations relations \eqref{commE} for the flux fields define a 2-sphere generalisation of an SU$(2)$ valued loop algebra. The representations are classified by the  SU$(2)^S$ whose Casimir is given by the corner area element $\rho(\sigma) = |\det(e_a^i)|(\sigma)$. 
At the classical level this object defines a measure on the sphere which is strictly positive.

At the quantum level we have to chose a representation of the operators we just described, that is we have to chose a unitary  representation of $G^S_{\mathrm{LQG}}$.
These representations are classified by the choice of sphere measure. 
We now focus our analysis to representations labelled by a  discrete measure on $S$ which now carry a new quantum number $N$: the number of punctures on the sphere where this measure is non-vanishing. 
 Representation states of SU$(2)^S$  are given by functions $\varphi: S^N \to \otimes_{i=1}^N V_{j_i}$ where $V_j$ denotes a spin-$j$ representations of SU$(2)$ and $N$ denotes the number of punctures activated on the corner sphere. On these states, the action of the flux operator is then simply given by 
\be \label{SU(2) flux action}
E_i(\sigma) \varphi(\sigma_1,\cdots\sigma_N) 
= \sum_{n=1}^N \delta^{(2)}(\sigma,\sigma_n) \rho_{j_n}(X_i) \varphi(\sigma_1,\cdots\sigma_N), 
\ee 
where $X_i$ is an SU$(2)$ generator and $\rho_j: \mathfrak{su}(2)\to V_j$ denote the spin-$j$ representations. This representation corresponds to a choice of discrete density  given by
\be\label{rhorep}
\hat{\rho}(\sigma)\varphi(\sigma_1,\cdots,\sigma_N)=\sum_{n=1}^N \rho_n \delta^{(2)}(\sigma,\sigma_n) \varphi(\sigma_1,\cdots,\sigma_N) ,
\ee 
where 
$
\rho_n = \gamma \hslash \sqrt{j_n(j_n+1)}
$
is the discrete area spectrum of LQG.

An interesting analogy comes by noticing that $G^S_{\mathrm{hydro}}=  \mathrm{Diff}(S)\ltimes\mathbb{R}^S$, which is the group generated by $(P_A, \rho)$, is isomorphic to the symmetry group of a 2-dimensional barotropic fluid\footnote{Barotropic fluids are such that the fluid pressure $P(\rho)$ is a function of the fluid density $\rho$ only. In general the pressure can also depend on the entropy density.} \cite{khesin1989invariants}.
In this analogy $\rho$ is the fluid particle density, and it is well-known that there exist two classes of fluid representations: the ones for which $\rho$ is a strictly positive continuous measure, and the ones for which $\rho$ is a discrete measure.
In the second case the fluid is composed of molecules. Here we see that the quantization of area is analog to having a corner quantum fluid made of fundamental quanta.

This analogy also allows us to classify the representations of Diff$(S)$. It is known in the mathematical physics litterature that these groups are non-anomalous \cite{Khesin} . This is why we can  assume that the quantum theory provides a representation for these groups.

As shown 
in \cite{Freidel:2021qpz}, the representation of the diffeomorphism symmetry group is then labelled by  conformal dimensions $(\Delta_1,\cdots,\Delta_N)$ and  spins $(s_1,\cdots,s_N)$. These representations are irreducible and also appears as representation labels for BMSW . The Casimirs associated with diffeomorphism have been constructed in \cite{Donnelly:2020xgu}. The action of the diffeomorphism generator $P_A$ on the discrete representation states is then given by 
\begin{align}\nonumber
\hat{P}_A(\sigma) \varphi(\sigma_1,\cdots,\sigma_N) =\; & \sum_{n=1}^N  \delta^{(2)}(\sigma,\sigma_i) \frac{\partial \varphi}{\partial \sigma_n^A}(\sigma_1,\cdots,\sigma_N)\\
&- \sum_{n=1}^N \big(\Delta_n \delta_A{}^B + s_n\epsilon_A{}^B\big)\frac{\partial \delta^{(2)} (\sigma,\sigma_n)}{\partial \sigma^B} \varphi(\sigma_1,\cdots,\sigma_N).\label{Conf}
\end{align} 
Here $\epsilon_{A}{}^B$ corresponds to the choice of a complex structure on the sphere, i.e. it satisfies $\epsilon_{A}{}^B\epsilon_{B}{}^C= -\delta_A{}^C$. We see that a generic discrete representation of the LQG corner symmetry group assigns 3 quantum numbers $(j,\Delta,s)$ to each puncture. The states associated with $(j,\Delta,s)$ can be expanded as a superposition of spin network states purely  labeled by SU$(2)$ indices. An explicit expansion is given in \cite{Freidel:2019ees}. 
It was shown in \cite{Wieland:2017cmf} that these discrete representations labeled by $N$ appear naturally as a Fock space quantization of the 2d corner symplectic fluid. 

These preliminary considerations should give ample motivations to explore the possibility of representing the corner symmetry group and connecting the ensuring representations to an extension of the quantum geometric picture arising from LQG, including in particular diffeomorphism symmetry.
An important point of clarification is that, although spin foam models do not represent diffeomorphism symmetry, canonical LQG does by simply allowing the embedded graph to be displaced by a diffeomorphism. What we can now clearly see from \eqref{Conf} is that the LQG representation of diffeomorphism is trivial, i.e. it correspond to choosing $(\Delta,s)=(0,0)$.  

\section{Boundaries in loop quantum gravity}
\label{sec:old}
We have seen in the previous section that corner symmetries allow us to understand the entanglement of quantum spacetime from the representation theory of corner symmetry groups. From the perspective of the corner symmetry groups, the fundamental quantum discreteness of  geometry is a simple consequence of representation theory. These results suggests a form of local holography, where the boundary data for local subregions plays a pivotal role in the construction of the physical quantum states.

These ideas are profoundly connected to earlier LQG insights, for example through the work on isolated horizons and black hole entropy \cite{BarberoG:2022ixy}, on the relationship between holography and quantum entanglement \cite{Bianchi:2023avf}, and on the geometrical understanding of spin networks and spin foams (e.g. through twisted geometries) \cite{Freidel:2010aq,Bianchi:2009tj,Freidel:2011ue}. One element that the new corner perspective brings however is the seamless connection with the continuum QFT, and the importance of diffeomorphism symmetry and its representation labels such as $(\Delta,s)$. It also opens up the possibility to connect with the vast literature on holographic approaches to quantum gravity, and provides new avenues of development  for the core LQG results such as a Lorentz covariant formulation of spin network states with discrete area spectra.

We now briefly summarize some results which highlight the role played by boundaries in LQG. We refer the reader to the other review chapters \cite{BarberoG:2022ixy} for black hole entropy, \cite{Bianchi:2023avf} for entanglement in LQG, \cite{Haggard:2023tnj} for further details on quantum geometry and spin networks, and \cite{Asante:2022dnj} for spin foams and renormalization.

\medskip
\noindent{\bf Black hole entropy in LQG.}
 One of the key ideas of LQG concerning black holes was proposed by Krasnov \cite{Krasnov:1996tb}, building up on previous work by Smolin \cite{Smolin:1995vq} about edge mode degrees of freedom in gravity and TQFTs. The key idea first developed was  that black hole entropy really counts the numbers of quantum degrees of freedom that live on the black hole horizon and represent quantum geometrical states \cite{Rovelli:1996dv, Ashtekar:1997yu}. Early on, a connection between these quantum states and horizon edge modes in the gravity phase space was conjectured \cite{Ashtekar:1999wa}. These results led on the classical side to the study of boundary conditions for black holes and the proposal of isolated horizons \cite{Hayward:1993wb,Ashtekar:2000eq,Ashtekar:2004cn}. It also led to a refined way of counting black hole microstates \cite{Domagala:2004jt,Meissner:2004ju,Agullo:2010zz}, and eventually to important realization of the connection with U$(1)$ and SU$(2)$ Chern--Simons theory living on the horizon \cite{Engle:2010kt,Engle:2011vf,Bodendorfer:2013jba,Bodendorfer:2013sja}. In these works, the central role of the boundary symplectic structure was put forward, and it was finally understood that horizon punctures could also support loop algebra symmetries \cite{Ghosh:2014rra}. These results from the black hole entropy counting led to a direct connection with the corner symmetry in \cite{Freidel:2016bxd}. Note also that the notion of isolated horizons was recently generalized to the notion of \emph{non-expanding horizons} \cite{Ashtekar:2021wld}. These horizons correspond to  null surfaces where the corner symmetry charges are conserved, and the corresponding Carrollian fluid is perfect, see section \ref{subsec:Carroll}.

\medskip
\noindent{\bf LQG and twisted geometries.}
 We have seen that the entangling sphere carries representations of the boundary symmetries. The traditional loop gravity picture assigns, on the other hand, geometrical data to the vertex inside the sphere. So there seems to be tension between the two interpretations. This tension is resolved by the twisted geometry interpretation of spin networks, developed in \cite{Freidel:2010aq,Freidel:2010bw}. In twisted geometries, one assigns geometrical elements to each link that intertwines two vertices and then proves that the matching of representations across the link allows one to reconstruct the connection. In a twisted geometry, there are two discrete geometries across the link: the one from the left and the one from the right. If one tries to identify the group data in terms of polyhedral geometries, one finds a mismatch: while the areas are matched, the shapes of the two  adjacent polyhedra might differ; hence the discrete geometry is \emph{twisted}. Remarkably, it was shown in \cite{Haggard:2012pm} that the difference in shapes can be encoded into an SL$(2,\mathbb{R})$ connection. In  \cite{Freidel:2018pvm} it was also shown, following \cite{Freidel:2015gpa}, that one has a natural generalization of spin networks states that also carries representations of  SL$(2,\mathbb{R})$ necessary to reconstruct the frame field. It was proven in \cite{Freidel:2020svx} that the generator of this  SL$(2,\mathbb{R})$ symmetry is, in the continuum, the induced geometry of the corner sphere transverse to the spin network link.  Finally, in \cite{Wieland:2021vef} the change of this SL$(2,\mathbb{R})$ was shown to encode radiation.

Demanding that the geometry is not twisted projects twisted geometries, which represent LQG states, onto Regge geometries \cite{Dittrich:2008ar}.
The fact that LQG geometries are twisted and differ from the Regge geometries creates a puzzle. This puzzle can be simply resolved if one accepts that the faces dual to loop gravity states are not planar, and the edges of the dual polyhedra are \emph{spinning} instead of being straight \cite{Freidel:2013bfa}.  Moreover, we know from \cite{Freidel:2011ue} that there is an isomorphism between twisted geometries representing LQG states and piecewise flat geometries which provides an exact continuum description of the LQG phase space in agreement with  the results  described in this review. This result also gives a geometrical understanding of the dual LQG vacua constructed in \cite{Dittrich:2014wpa,Dittrich:2014wda,Dittrich:2016typ}.  Another prescient paper by Bianchi \cite{Bianchi:2009tj}  proposed early on a similar understanding of classical LQG geometry compatible with the action of diffeomorphisms. Now is a good time to go back to these results and use them to connect LQG and corner symmetries more tightly. 

\medskip
\noindent{\bf  Entanglement in LQG, spin foams and group field theories.}  
 Very early on it was  proposed that the area entanglement and its relation to boost symmetry could be recovered from the spin network entanglement of links carrying SL$(2,\mathbb{C})$ representations \cite{Bianchi:2012ui,Bianchi:2012vp}. Since then 
there has been a recent interest in computing entanglement entropy of spin network states and trying to relate it to the Ryu--Takayanagi formula \cite{ Chirco:2021chk,Han:2016xmb}.
The central idea here is that quantum geometry arises from a \emph{network of entanglement}. The quantum entanglement in spin networks is encoded through the representation link \cite{Donnelly:2008vx, Donnelly:2011hn,Livine:2017fgq,Baytas:2018wjd}. Similar ideas about entanglement have been explored in GFTs \cite{Chirco:2021chk}.
These results are a discrete expression of the fundamental ideas presented earlier, namely that entanglement can be derived from  corner symmetry. It would be really interesting to develop further the connection between entropy counting and the renewed understanding of symmetry. In particular, in order to connect to the Unruh calculation, one would need to construct, in quantum gravity, the  spacetime boost operator.  Again, we refer the reader to the chapter \cite{Bianchi:2023avf} for more details about entanglement in LQG.

\medskip
\noindent{\bf Spin foams as quantization of boundary data.} 
 Spin foams are based on a lattice truncation. The basic idea is to glue flat building blocks in such a way that Einstein's equations are satisfied at large scales compared to the typical lattice scale. The resulting spin foam amplitudes are a sum over geometric data in each fundamental four-simplex \cite{Perez:2012wv,Wieland:2016exy,Asante:2022dnj}. Each four-simplex is flat and contains no radiative data inside. It is only through the non-trivial gluing between adjacent four-simplices that curved geometries arise. Curvature is distributional and concentrated at surfaces dual to the fundamental triangles.  In fact, it goes deeper than this: the traditional formulation of spin foam models focuses on simplicity constraints at the vertices of the spin foam \cite{Engle:2007wy, Freidel:2007py}, thereby restricting the geometry of boundary tetrahedra. One can take an equivalent point of view where the simplicity constraints are imposed by selecting appropriately the propagator going between the spin foam vertices, hence by controlling the corner geometry matching. This point of view adapted to corner and twisted geometry is already present in \cite{Freidel:2007py}, it was further developed in \cite{ Kaminski:2009fm} and shown to be extremely efficient in the proof of spin foam asymptotics \cite{Banburski:2014cwa}. This twisted geometry perspective on spin foams was first discovered in the context of GFT models for quantum gravity \cite{Freidel:2005qe,Oriti:2009nd} where one witnessed that it is possible to encode quantum gravity constraints by the right choice of propagator while keeping the interaction vertex simple. 

\medskip
\noindent{\bf What is missing?}
In asymptotically flat spacetimes, the simplest Dirac observables are the ADM energy and momentum, angular momentum and centre of mass together with an infinite-dimensional algebra of corner charges. The same happens at finite distance. There are infinitely many corner charges that are Dirac observables on the physical phase space of a bounded region. These  charges generate corner symmetries such as super-rotations (which represent  tangent diffeomorphisms), super-boosts and super-translations. We have seen in section \ref{sec:quantum geo} that, even when we are  in a vacuum state,  some of the charges such as the corner boost charges do not vanish. Therefore, a generic physical state should carry a non-trivial representation of the corner symmetries. The problem with the usual spin network representation is that the basis states do not carry the corresponding representation labels. 
We know that these labels characterize components of the 4d metric at the corner. For instance, the generator of Diff$(S)$ symmetry knows about the off-diagonal components of the metric in metric gravity \cite{Donnelly:2016auv} or knows about the torsion of the Ashtekar--Barbero connection in loop gravity \cite{Freidel:2015gpa,Freidel:2016bxd, Freidel:2019ees}. Another example concerns the  geometry  symmetry encoded into an SL$(2,\mathbb{R})$ edge group, as  described in the next section. Its representation labels describe quantization of the tangential metric component \cite{Freidel:2015gpa, Freidel:2020svx}.  Importantly, we  lack a proper representation of asymptotically-flat spacetimes in terms of spin network states (see \cite{Campiglia:2014xja}) which carries a representation of BMS symmetry. By connecting the boundary charges in a finite region to asymptotic charges at infinity, we obtain a new perspective to this long-standing problem in the field (see \cite{Freidel:2021cjp,Wieland:2020gno, Campiglia:2020qvc} for this connection at the semi-classical level).

By not representing these symmetries at the quantum level, we lose information about the quantum geometry. 
In LQG, we have quantum numbers for internal SU$(2)$ frame rotations  and corresponding intertwiners, but there are no representation labels for the rest of the corner symmetry group. For instance as shown in \cite{Freidel:2016bxd}, with SU$(2)$ labels we can represent the flux but miss SL$(2,\mathbb{R})$ labels needed to represent the frame.

In \cite{Freidel:2016bxd} we have shown preliminary results on how to include the conformal weights and spin labels of Diff$(S)$ representations, see   \eqref{Conf}, in the spin network description. The main result is that these edge labels  control the gluing of states when we glue two elementary quantum subregions along an edge. It forces specific spin network superpositions to enter, such as $|\Delta,s\rangle\langle \Delta, s| =\sum_{j\geq s} \rho_j(\Delta,s) |j,m\rangle\langle j,m|$. These superpositions arise from the matching of diffeomorphism symmetry which determines the weights $\rho_j(\Delta,s)$. Therefore, having extra representation labels does not necessarily mean that spin networks are a wrong basis. It means that no physical states can be supported on a single spin network state. Since the nature of quantum geometry entanglement cannot simply be described by sharing an SU$(2)$ spin. One needs extra symmetry labels control what superposition of states is allowed when we glue two quantum subregions together. In other words, the quantum geometry entanglement of subregions is considerably reinforced by the presence of  diffeomorphism symmetry at the quantum level\footnote{As we have seen in section \ref{sec:representation}. The usual formulation of LQG contains only trivial representations of the diffeomorphism group that do not create any new entanglement.}.  Ultimately this is what we want: the demand of invariance under super-rotation, super-boost and super-translation should constrain the admissible spin network superposition allowed in the edge gluing by requiring that these form representation states of the corner symmetry group. The fact that after gluing the representation labels are matched is the expression that the Hamiltonian and diffeomorphism constraints are implemented.  
So it should be now clear that not including these representation labels in the construction of spin network states in LQG is problematic. It is therefore of utmost importance to revisit the investigations of quantum geometry   entanglement described in \cite{Bianchi:2023avf} in light of these results. 

For the spin foam models, we can still use boundary states which are spin networks. What the corner symmetry analysis provides, however, is the possibility to constrain the value of spin foam amplitudes via symmetries. It is well-known that in the presence of symmetries, path integral amplitudes satisfy Ward identities which express the invariance of the quantum amplitude under a symmetry transformation of the boundary state. We should therefore expect spin foam models and GFT models to satisfy such Ward identities. These identities represent the quantization of the infinite set of flux-balance laws that arise from solving the Hamilton and vector constraints.
At asymptotic infinity, these balance laws turn into Ward identities, called soft theorems, that put stringent constraints on the perturbative $S$-matrix \cite{Strominger:2017zoo}. 
This means that a new exciting possibility for spin foam models is opening up: the possibility to identify the quasi-local generators of symmetry charges acting on the spin network states and write down the Ward identities that spin foam models ought to satisfy.
\subsection{Classical theory and symplectic structure}
\label{sec:new}

 We have already briefly discussed above in section \ref{sec:representation} how
 LQG representation of quantum geometry naturally arises from studying the corner symmetries group appearing in tetrad gravity with a gauge-fixed internal normal. The choice of this internal normal  breaks Lorentz invariance and  determines an internal SU$(2)$ group.
 Our goal is now to show how a \emph{generalization} of the LQG representation of quantum geometry, covariant under the internal Lorentz group, naturally arises  
  from studying the corner symmetries of tetrad gravity. This result follows from a detailed canonical analysis 
 of the first-order Lagrangian for gravity.

Let us consider a tetrad 1-form $e^I$ and a connection 1-form $\omega^{IJ}$ with curvature $F^{IJ}$. These fields can be used to build the Lagrangian 4-form for Einstein--Cartan--Holst gravity
\be\label{ECH Lagrangian}
L_\text{ECH}=\f{1}{2}E_{IJ}\wedge F^{IJ},
\q\q
E_{IJ}[e]=(\star+\beta)(e_I\wedge e_J).
\ee
We are going to focus primarily on the symplectic structure of this Lagrangian. In its analysis, a crucial role is played by the internal normal $n^I=n^\mu e^I_\mu$ obtained from the time-like normal $n^\mu$ to a spacelike hypersurface $\Sigma$. This normal is such that $n^In_I=n^\mu n_\mu=-1$. The importance of keeping track of this normal in the phase space was already recognized in various contexts in \cite{Bianchi:2012vp,Bodendorfer:2013hla,Alexandrov:2002br,Wieland:2010ec,Wieland:2014nka,Wieland:2017zkf,Bodendorfer:2011nw,Bodendorfer:2013jba}.

Using the internal normal, we can decompose the Lorentz connection as
\be
\omega^{IJ}=\Gamma^{IJ}-2K^{[I}n^{J]},
\q
\de_\Gamma n^I=0,
\q
\de_\omega n^I=K^I.
\ee
We also decompose the pull-back of the tensorial 2-form $E^{IJ}$ as
\be
E^{IJ}|_\Sigma=-2E^{[I}n^{J]}+\beta(e^I\wedge e^J).
\ee
As explained at length in \cite{Freidel:2020svx}, with these decomposition the symplectic structure of the ECH Lagrangian becomes
\be\label{ECH structure}
\Omega_\text{ECH}=\int_\Sigma\delta K^I\wedge\delta E_I+\oint_S\left(\delta E_I\delta n^I-\f{\beta}{2}\delta e_I\wedge\delta e^I\right).
\ee
One can see that the normal appears in the boundary contribution, and also that the frame is conjugated to itself on the boundary due to the non-vanishing Barbero--Immirzi parameter. To interpret this result, one should notice that the bulk term is nothing but the symplectic structure of canonical gravity. Indeed, with a slight rewriting, one can show that \cite{Freidel:2020svx}
\be
\int_\Sigma\delta K^I\wedge\delta E_I=\int_\Sigma\delta(K\times e)^I\wedge\delta e_I=\f{1}{2}\int_\Sigma\delta p^{ab}\delta h_{ab}=\Omega_\text{GR},
\ee
where $p^{ab}=\sqrt{h}\big(Kh^{ab}-K^{ab}\big)$ is the usual momentum density of the ADM formulation. In summary, we can write \eqref{ECH structure} in the form
\be
\Omega_\text{ECH}=\Omega_\text{GR}+\Omega^S_\text{ECH/GR},
\ee
where $\Omega^S_\text{ECH/GR}$, which is the boundary contribution in \eqref{ECH structure}, is the relative boundary symplectic structure between the ECH and canonical formulations of gravity. This is the main message of \cite{Freidel:2020xyx}, namely that any formulation $L_\text{F}$ of gravity has a symplectic structure which is of the form $\Omega_\text{F}=\Omega_\text{GR}+\Omega^S_\text{F/GR}$. The bulk symplectic structure is universal and encodes the fact that the theories being discussed are general relativity, while the boundary contribution is formulation-dependent. Even if ECH gravity \eqref{ECH Lagrangian} is equivalent in the bulk to general relativity, it carries a specific boundary symplectic structure.

Our goal is to understand the physical significance of the boundary symplectic structure \eqref{ECH structure}. This latter is responsible for the appearance of non-vanishing charges for Lorentz transformations, and, as we will explain shortly, for the non-commutativity of the tangential boundary metric (which in turn leads to the discreteness of the boundary area). Note that studies of the relation between the Einstein--Hilbert and Einstein--Cartan--Holst formulations through a boundary term in the symplectic structure go back to \cite{Peldan:1993hi}. The emphasis on relative charges was made more explicit in \cite{DePaoli:2018erh,Oliveri:2019gvm}.

\subsection{Charges and discreteness}

First, let us stress that the treatment presented here is in a sense orthogonal to the usual approach to canonical LQG. Indeed, in standard treatments of LQG one focuses on bulk variables, and in particular the contribution of the Holst term proportional to $\beta$ is used to build a bulk connection variable. This is made possible by the identity $\beta\de(e_I\wedge\delta e^I)\hat{\;=\;}\beta(e_I\wedge e_J)\wedge\delta\omega^{IJ}[e]$, which holds on-shell of the torsion equation of motion. In the time gauge where $n^I=\delta^I_0$, this reduces to $\beta\de(e_i\wedge\delta e^i)\hat{\;=\;}\beta E^a_i\delta\Gamma^i_a[e]$, where $i$ is the index in the $\mathfrak{su}(2)$ subalgebra of $\mathfrak{so}(3,1)$ which survives the fixing of the normal. This is the identity that enables, once the Holst term is included, to define a bulk Ashtekar--Barbero connection variable conjugated to the electric field $E$. Here we wish to shift emphasis from the bulk to the boundary, and therefore keep the contribution of the Holst term in the boundary symplectic structure.

The fact that the symplectic structure of ECH gravity differs from that of canonical gravity $\Omega_\text{GR}$ by the presence of a surface term has the important consequence that it gives rise to a new set of surface charges. These are the charges of the internal Lorentz transformations acting as $\delta_\alpha V^I=-{\alpha^I}_JV^J$ for $V^I=(E^I,n^I,e^I)$ and $\delta_\alpha\omega^{IJ}=\de_\omega\alpha^{IJ}$. These charges are obtained as usual by computing the contraction $I_\alpha\Omega$, and their explicit expression is
\be
G(\alpha)=\int_\Sigma E^{IJ}\wedge\de_\omega\alpha_{IJ}\hat{\;=\;}\oint_S\alpha_{IJ}E^{IJ}=\oint_S\alpha_{IJ}\Big(\beta(e^I\wedge e^J)-2E^In^J\Big).
\ee
These boundary charges are the  Lorentz covariant version of the SU$(2)$ charges \eqref{LQG SU2 charges}. They have also been discussed extensively in the context of black hole entropy counting in \cite{Bodendorfer:2013sja}. It is important to stress out that these charges are purely relative charges, in the sense that they are produced entirely by the boundary symplectic structure in \eqref{ECH structure}. This means that, as expected, metric gravity with the symplectic structure $\Omega_\text{GR}$ does not possess Lorentz charges. This is the key property which differentiates, already at the classical level, LQG from any metric based approach: it possesses extra surface charges which give rise to the basic surface flux operators. We have here presented the covariant formulation with an arbitrary internal normal $n^I$, but in LQG the normal is fixed to $\delta^I_0$ and one recovers the SU$(2)$ fluxes \eqref{LQG SU2 charges} whose action on states is \eqref{SU(2) flux action}.

We now turn to the main result obtained from the detailed analysis of the boundary symplectic structure, which is the discreteness of area implied by the presence of $\beta\neq0$. In \cite{Freidel:2020ayo,Freidel:2020svx} we have decomposed the 6 components\footnote{There are 6 components because $e^I_an_I=0$ and on $S$ we have $a=1,2$.} of the tangential frame $e^I_a$ at the surface $S$ into a so-called spin operator $S^I=\f{1}{2}\beta(e\times e)^I$ and the tangential metric $q_{ab}=e^I_ae^J_b\eta_{IJ}$. The spin operator is the SU$(2)$ flux $E^i$ written in internal Lorentz-covariant form. Equipped with this decomposition, one can show that the tangential metric satisfies the $\mathfrak{sl}(2,\mathbb{R})^S$ algebra
\be
\lb q_{ab}(x),q_{cd}(y)\rb=-\f{1}{\beta}\big(q_{ac}\eps_{bd}+q_{bc}\eps_{ad}+q_{ad}\eps_{bc}+q_{bd}\eps_{ac}\big)(x)\delta^2(x-y).
\ee
This is in line with earlier analysis carried out in \cite{Freidel:2018pvm,Freidel:2015gpa} in the SU$(2)$ gauge fixed formulation. This important result shows that the corner metric becomes non-commutative in the presence of the Barbero--Immirzi parameter! This shows the tangential metric component of $S$ are the $\mathfrak{sl}(2,\mathbb{R})^S$ generators.
One can then show that the quadratic Casimir is given by
\be
C_{\mathfrak{sl}(2,\mathbb{R})^S}=\beta^2\det(q).
\ee
Since the metric is that of a 2-dimensional space-like surface, the determinant on the right-hand side is positive, and we conclude from this relation that the $\mathfrak{sl}(2,\mathbb{R})^S$ Casimir is positive as well. This indicates that the surface area spectrum has to be labeled by the discrete series of representations, if we were to label states by unitary representations respecting the boundary symmetry structure.

Note that in order to make this construction rigorous a regularization of the corner algebra is necessary, as outlined in \cite{Freidel:2018pvm}. A treatment that also includes diffeomorphism symmetry on the sphere is also required, along the lines of \cite{Freidel:2019ees,Freidel:2016bxd,Donnelly:2020xgu,Wieland:2017cmf,Donnelly:2022kfs}. In particular, in \cite{Wieland:2017cmf} a regularisation in terms of spherical harmonics with respect to a fiducial metric at the sphere was introduced as an example of a regularisation which does not involve any discretization. In \cite{Donnelly:2022kfs} such a regularization was shown to be equivalent to a matrix model regularization and arising by rendering the corner sphere non-commutative. This is ultimately related to the understanding of quantization of diffeomorphisms in the corner symmetry group and the role of the dynamics.  We therefore now turn to a description of the dynamics along null boundaries and show that in this context we can also obtain the discreteness of area  from the canonical analysis of null surfaces.

\section{Null boundaries, isolated horizons and fluid conservation law}
\label{sec:null}

In the previous sections, we have discussed the representation of the corner symmetry group which contains the symmetry generators with vanishing flux. In order to understand the dynamics one needs to describe the representation of the extended corner symmetry group and include the super-translation generators that move the corner transversally. It has proven invaluable to focus on the representation of null super-translations and study the gravitational dynamics projected along null surfaces and null horizons. There has been a rich literature of the subject dating back to the membrane paradigm \cite{price1986membrane}, the study of isolated horizons \cite{Ashtekar:1999wa}, and more recently the dynamics of null surfaces, the construction of the symplectic potential, of the symmetry charges, and their understanding in terms of Carrollian geometry \cite{Donnay:2016ejv, Hopfmuller:2016scf, Hopfmuller:2018fni, Chandrasekaran:2018aop, Donnay:2019jiz, Wieland:2020gno, Chandrasekaran:2021hxc, Adami:2021nnf, Freidel:2022vjq}.

\subsection{Carrollian fluid conservation law}
\label{subsec:Carroll}

In this section we study the geometry and dynamics of a null boundary $\mathcal{N}$. On a null boundary the pull-back of the four-dimensional metric is degenerate. The degenerate direction is the direction of the null rays that generate the surface. This null direction is in the kernel of the induced metric
\begin{equation}
q_{ab}=\varphi^\ast_{\mathcal{N}} g_{ab},\qquad \ell^aq_{ab}=0.
\end{equation}
Any such null direction is unique up to a local rescaling sending\footnote{We assume that $\ell^a$ is future pointing.} $\ell^a$ into $\mathrm{e}^\lambda\ell^a$. Mathematically this means that $\mathcal{N}$ is a fibered manifold with  fibration $p: \mathcal{N} \to S $. The null direction $\ell$ is in the kernel of $\de p$ (push-forward of $p: \mathcal{N} \to S$) while the null metric $q_{ab}$ can be understood as  the pull-back of a time dependent metric $q^S_{AB}$ on the base. The Lie derivative of the metric along $\ell $ decomposes in terms of the expansion and shear of the null surface:
\be 
\theta_{ab}\coloneqq  \frac12 {\cal L}_\ell q_{ab}= \frac{q_{ab}}{2} \theta + \sigma_{ab}. 
\ee 
Note that the degenerate metric $q_{ab}$ determines a spatial area form $\epsilon_S$ on $\mathcal{N}$ such that $i_{\ell} \epsilon_S=0$. Moreover the choice of a null generator $\ell$ determine a volume form $\epsilon_{\mathcal{N}}$ on $\mathcal{N}$ which is such that $i_{\ell}\epsilon_{\mathcal{N}}= \epsilon_{S} $.
The expansion relates the two forms through the identity $ \de\epsilon_{S}=\theta \epsilon_{\mathcal{N}}$.

In order to construct a connection on $\mathcal{N}$ and  describe the dynamics of null surfaces, it is convenient to introduce a Ehresmann connection. This is a 1-form $k_a$ dual to the null vector $\ell^a:k_a\ell^a=1$ \cite{Bekaert:2015xua, Ciambelli:2019lap}. The Ehresmann connection defines a notion of horizontality, where $Y$ is an horizontal vector field on $\mathcal{N}$ if $i_Y k=0$. A general vector $\xi \in T\mathcal{N}$ can be decomposed as $\xi=T\ell + Y$, with $Y$ horizontal. The Ehresmann connection also allows to decompose the volume form on $\mathcal{N}$ as $\epsilon_{\mathcal{N}}=k\wedge \epsilon_S $. 
The data ${\mathcal C}_{\mathcal{N}}=(q_{ab}, \ell^a, k_b)$ represents the \emph{ Carrollian geometry} of $\mathcal{N}$.
This data defines a \emph{rigging structure} \cite{Mars:1993mj}, i.e. a projector  $\Pi_a{}^b= q_a{}^b + k_a\ell^b$ where $q_{a}{}^b$ is the horizontal projector $\ell^a q_a{}^b=0=q_a{}^b k_b$. The presence of a rigging structure allow us to uniquely raise indices of horizontal tensors. For instance given the expansion tensor $\theta_{ab}$, the tensor $\theta_a{}^b$ is defined as the unique tensor such that $\theta_a{}^c q_{cb}=\theta_{ab}$ and such that $\theta_a{}^b k_b=0$.

The rigging structure allows us to introduce the notion of  \emph{Carrollian} connections $D_a$ \cite{Chandrasekaran:2021hxc,  Freidel:2022vjq}. These are torsionless connections that preserve the rigging projector: $D_a \Pi_b{}^c =0$.
A Carrollian connection defines a 1-form $\omega_a$ on $\mathcal{N}$ given by the derivative of the volume form:
\be 
D_a \epsilon_{\mathcal{N}} = -\omega_a \epsilon_{\mathcal{N}}.
\ee
This 1-form can be decomposed into a transverse and horizontal component as $ \omega_a =\kappa k_a + \pi_a$, where $\pi_a$ is the Hajicek connection, while $\kappa$ is the surface gravity measuring the inaffinity of $\ell$. $\omega_a$ can be understood as a boost connection since, under the boost rescaling  $\ell\to e^\lambda \ell$ and $k\to e^{-\lambda} k$, it transforms as $\omega_a \to \omega_a + \partial_a \lambda$.  One can then prove that the boost covariantized derivative of $\ell^b$ and $k_b$ are horizontal. They define the expansion and transverse expansion tensors
\be 
(D_a-\omega_a)\ell^b =\theta_a{}^b, \qquad 
q_a{}^c(D_c+\omega_c)k_b = {\theta}^{(k)}_{ab}.
\ee 
The Carrollian connection preserves the metric when derivatives and tensor indices are taken to be transverse directions. In general, one has 
\be 
D_a q_{bc}= -(k_b \theta_{ac}+ k_c \theta_{ac}).
\ee
We therefore see that the Carrollian connection compatible with the Carroll structure  $(q_{ab},\ell^b,k_b)$ is determined by the boost 1-form and the transverse expansion $(\omega_a,\theta^{(k)}_{ab})$.
It is well-known that the projection of Einstein's equation along the null surface leads to two sets of equations: the Raychaudhuri  and Damour equations \cite{damour1978black}. These are given by 
\be
(\ell+\theta)[\theta] &= \mu \theta -\sigma_{ab}\sigma^{ab},\\ 
q_a{}^b \left({\cal L}_\ell+\theta\right)[\pi_b] &= \overline{D}_a\mu - \overline{D}_b\sigma^{a}{}^b,
\ee
where we have introduced the surface tension $\mu \coloneqq  \left(\kappa +\frac{\theta}{2}\right)$ of $\mathcal{N}$ and $\overline{D}_a=q_a{}^b D_b$ denotes the horizontal derivative.

As shown in \cite{Hopfmuller:2018fni}, these equations can be understood as conservation equations of corner symmetry charges associated with super-translations along $S$ and diffeomorphism along $S$. The corresponding charge aspects are $\theta \epsilon_S$ for super-translations and $\pi_A\epsilon_S$ for super-rotations. These two equations can be understood as conservation  equations for Noether charges with non-trivial flux as in \eqref{Charge-Flux}, where the symplectic potential\footnote{This expression is valid in the boost frame where $k_a\delta \ell^a=0$. The boost transformation is pure gauge in the metric formulation of gravity while it possesses a non-trivial charge in the first order formulation as we are about to see.} on $\mathcal{N}$ is given by \cite{Parattu:2015gga, Hopfmuller:2016scf, Chandrasekaran:2020wwn}
\be 
\Theta_{\mathcal{N}} =\int_{\mathcal{N}} 
\left( \frac12 \big(\sigma^{ab} -\mu q^{ab}\big)\delta q_{ab} -\pi_a \delta \ell^a \right)\epsilon_{\mathcal{N}} .
\ee
From this we can read the canonical conjugate pairs of spin-0 $(\mu,\sqrt{q})$, spin-1 $(\pi_a, \ell^a)$ and spin-2 $(\sigma^{ab}, q_{ab})$.

Quite remarkably, these equations are now understood as conservation equations for a Carrollian fluid \cite{Donnay:2019jiz, Chandrasekaran:2021hxc, Freidel:2022vjq}. The Carrollian fluid stress tensor is given in terms of the rigging structure and the Carrollian connection simply as 
\be 
T_a{}^b \coloneqq D_a\ell^b - \Pi_a{}^b D_c\ell^c.
\ee 
The tensor $D_a\ell^b =\omega_a \ell^b +\theta_a{}^b$ is called the shape operator or Weingarten map.
As first shown in \cite{Chandrasekaran:2021hxc} (see also \cite{Freidel:2022vjq}) the Einstein equations pulled back on $\mathcal{N}$ can then simply be written as fluid conservation equations
\be 
\Pi_a{}^b G_{\ell b} = D_b T_a{}^b=0.
\ee 
This generalizes for null surfaces the celebrated Brow--York result \cite{brown1993quasilocal}.
In \cite{Freidel:2022vjq} it was shown that these equations can be obtained from the gravitational symplectic potential and that the symmetry charges associated with a vector field $\xi$ tangent to $\mathcal{N}$ are 
\be 
Q(\xi)= \int_S \xi^aT_a{}^b \epsilon_b,
\ee 
where $\epsilon_b =i_{\partial_b} \epsilon_{\mathcal{N}}$.
The Carrollian fluid energy-momentum tensor can be decomposed in terms of the quantities defined above as
\be 
T_a{}^b = -\theta k_a\ell^b + 
\pi_a \ell^b  - \mu  q_a{}^b +
\sigma_a{}^b.
\ee 
This energy-momentum tensor can be interpreted as that of a Carrollian fluid obtained from the $c\to 0$ limit of a relativistic energy-momentum tensor. In this fluid analogy, $\theta$ plays the role of the fluid energy, $\pi_a$ that of the heat flux, $-\mu$ of the pressure and $\sigma_{ab}$ of the viscous stress tensor.
We therefore see that it is not only the representation theory of the corner symmetry group that leads to a 2-dimensional fluid analogy, but also the gravitational dynamics along null surfaces in agreement with the membrane paradigm \cite{price1986membrane}. The notion of non-expanding horizons \cite{Ashtekar:2021wld, Ashtekar:2021kqj} can now be simply understood in this modern language as a perfect fluid where the energy $\theta$ and the dissipation tensor $\sigma_{ab}$ vanishes.

In order to connect these results to the LQG literature we now translate the analysis of  dynamics on a null surface in terms of tetrads and connection variables. This will allow us to understand the role of the Immirzi parameter and show that a non-zero Immirzi parameter implies a discretization of the spectra of the area operator which is the corner charge of internal boost symmetry. We also show that the spinor variables are naturally quantized as Fock space variables. These results are the null slice representation of the results derived in section \ref{sec:representation}  and \ref{sec:new}.

\subsection{Boundary frame fields and boundary Lagrangian}

Loop gravity is based on Einstein--Cartan geometry. Just like we have triads and tetrads, we can also have orthonormal co-frames intrinsic to the null boundary. The intrinsic metric is positive semi-definite and has one degenerate null direction. We introduce co-dyads $m_a$ and $\bar{m}_a\in\Omega^{1}(\mathcal{N}:\mathbb{C})$ that diagonalise the signature $(0$$+$$+)$ metric of the null surface
\begin{equation}
q_{ab}= 2m_{(a}\bar{m}_{b)}.\label{qmdef}
\end{equation}
Given $q_{ab}$, the co-dyad $(m_a,\bar{m}_a)$ is unique up to a local U$(1)$ transformation sending $m_a$ into $\mathrm{e}^{\mathrm{i}\varphi}m_a$. This transformation will play an important role below. Besides the metric $q_{ab}$, there is also a canonical 2-form at the boundary, namely the area-element
\begin{equation}
\varepsilon_{ab}=-2\mathrm{i}\,m_{[a}\bar{m}_{b]}.
\end{equation}
The oriented area of any 2-dimensional cross-section $S$ of $\mathcal{N}$ is then given by the integral $\mathrm{Ar}[S] = -\mathrm{i}\int_{S}m\wedge\bar{m}$.


 The loop representation is based on quantum states that carry half-integer spin labels. Rather than working with Lorentz vector-valued $p$-forms, it is therefore more natural to work with a spinor representation of the frame bundle. Using the soldering forms $\sigma^{AA'}{}_{I}$ between Lorentz vectors and spinors, we map Lorentz vectors into pairs of spinors,
\begin{equation}
V^{AA'}=\frac{\mathrm{i}}{\sqrt{2}}\sigma^{AA'}{}_I V^I,\qquad V^I=\frac{\mathrm{i}}{\sqrt{2}}\bar{\sigma}_{AA'}{}^IV^{AA'}.
\end{equation}
Under this map, the Plebański 2-form  $E^{IJ}=e^I\wedge e^J$ and the SO$(1,3)$ connection $\omega^{IJ}$ split into left-handed and right-handed parts.
We can now move to the construction of the boundary Lagrangian.

In the spinor representation, the action neatly splits into left-handed and right-handed parts,
\begin{equation}
S_{M}[A,e]=\left[\frac{\mathrm{i}}{8\pi\gamma G}(\gamma+\mathrm{i})\int_{M}\Sigma_{AB}\wedge F^{AB}\right]+\mathrm{cc}.
\end{equation}
We now need to explain how to couple this action to a boundary that is null. First of all we note that the pull-back of the self-dual 2-form $\Sigma_{AB}$ onto the null surface 
\begin{equation}
\varphi_{\mathcal{N}}^\ast\Sigma_{AB} = e_{(A}\wedge\bar{m}\,\ell_{B)}.\label{bndryvardef}
\end{equation}
These boundary fields have a neat geometric interpretation. The spinor $\ell_A$ is the square root of the null generators of $\mathcal{N}$. If $e_{AA'}$ is the bulk tetrad, we have, in fact,
\begin{equation}
\varphi^\ast_{\mathcal{N}}\ell_a=0,\quad \ell_a=\mathrm{i}e_{AA'a}\ell^A\bar{\ell}^{A'}.
\end{equation}
The  2-form $e_A\wedge\bar{m}$, on the other hand, encodes the rest of the intrinsic geometry of $\mathcal{N}$. If we contract, for example, $e_A\wedge\bar{m}$ with $\ell^A$, we obtain the area 2-form,
\begin{equation}
\varepsilon = \mathrm{i}\,e_A\wedge\bar{m}\,\ell^A.\label{epsdef}
\end{equation}
It is possible to show, see \cite{Wieland:2017zkf}, that one can recover the entire intrinsic geometry of the null surface from $\eta_A = e_A\wedge\bar{m}$ and $\ell^A$ alone.

To introduce a basis in the spin bundle, we introduce a dual spinor $k_A$, whose square returns the Ehresmann connection introduced earlier in section \ref{subsec:Carroll}, i.e.\
\begin{equation}
k_a = -\mathrm{i}\,e^{AA'}{}_{a} k_A \bar{k}_{A'},\quad\text{such that}\quad k_A\ell^A =1.
\end{equation}
In terms of this basis, the spinor-valued 2-form $e_A\wedge\bar{m}$ admits the decomposition
\begin{equation}
e_A\wedge\bar{m} = -(\ell_A k+k_A m)\wedge{\bar{m}},\label{etadef}
\end{equation}
where $k$ is the Ehresmann connection on the null boundary (now viewed as a 1-form intrinsic to $\mathcal{N}$). The curvature of this connection determines the Carrollian acceleration $\varphi$ and the Carrollian vorticity $w$ \cite{Ciambelli:2019lap,Freidel:2022vjq}.
\begin{equation}
\mathrm{d} k = -\varphi k\wedge\bar{m}-\bar{\varphi}k\wedge m+\mathrm{i}\,w\,m\wedge\bar{m}.
\end{equation}
The exterior derivative of the U$(1)$ dyad $m_a$, on the other hand, determines a U$(1)$ connection $\Gamma$ and the shear $\sigma $ and expansion $\theta$ of the null boundary via
\begin{equation}
\mathrm{d}m =-\mathrm{i}\,\Gamma\wedge m +\frac{1}{2}\theta k\wedge m+\sigma k\wedge \bar{m}.
\end{equation}
Taking into account the Frobenius theorem, we thus see that for non-vanishing shear, there is an obstruction to find a potential for $m_a$. If there is shear, we can not find a holomorphic coordinate $z:\mathcal{N}\rightarrow\mathbb{C}$ that would satisfy $m\propto\mathrm{d} z $.

Derivatives of the boundary spinors are important as well. Without assuming special boundary conditions, the only available derivative  is the pull-back of the SL$(2,\mathbb{C})$ covariant (exterior) derivative from the bulk. We call this derivative $D=\varphi^\ast_{\mathcal{N}}\nabla$ and it provides a Carrollian connection as explained in the previous section. Given the spinor basis $(k^A,\ell^A)$ at the null boundary, the spin coefficients are
\begin{align}
k_A D\ell^A & = \frac{1}{2\mathrm{i}}\Big(\Gamma+\mathrm{i} \omega \Big),\\
\ell_A D\ell^A &= -\Big(\frac{1}{2}\theta m+\sigma \bar{m}\Big),\\
k_A Dk^A & = -\frac{1}{2}\big(\theta_{({k})}-\mathrm{i}\,w\big)\bar{m}-\bar{\sigma}_{(k)}\bar{m}+\big(\bar{\varphi}+\bar{\pi}\big)k,
\end{align}
where the 1-form $\omega\in\Omega^1(\mathcal{N}:\mathbb{R})$ encodes the non-affinity $\kappa$ of the null generators $\ell^a$ as well as the momentum aspect $\pi_a$ of the Carrollian fluid, i.e.\ 
\begin{equation}
\omega =  \kappa k +\bar{\pi}m+\pi\bar{m}.
\end{equation}
The 1-form $\omega\in T^\ast \mathcal{N}$ determines extrinsic data---it can not be obtained from exterior derivatives of $m$ and $k$ intrinsic to $\mathcal{N}$. In the same way, the transversal expansion $\bar{\theta}=q^{ab}\nabla_ak_b$ and the transversal shear $\sigma_{(k)}=\bar{m}^a\bar{m}^b\nabla_a k_b$ are extrinsic quantities on $\mathcal{N}$.\footnote{In here, we extended the boundary fields $\ell^a\in T\mathcal{N}$ and $(k_a,m_a,\bar{m}_a)\in T^\ast\mathcal{N}$ into a Newman--Penrose null tetrad  $(k^a,\ell^a,m^a,\bar{m}^a)$ in a neighbourhood of $\mathcal{N}$ such that $q^{ab}=m^a\bar{m}^a+\mathrm{cc}.$} In the following, it will be useful to combine $\Gamma$ and $\omega$ into a single complexified boundary connection $A\in \Omega^1(\mathcal{N}:\mathbb{C})$, which is given by
\begin{equation} 
A = -\mathrm{i}\,\Gamma+\omega-\theta k.\label{Adef}
\end{equation}

The boundary fields carry a representation of the boundary symmetries. First of all, they are clearly covariant under internal SL$(2,\mathbb{C})$ gauge transformations and diffeomorphisms that preserve $\mathcal{N}$. This is trivially so. Then they also admit a complexified U$(1)$ symmetry that acts on boundary fields
$(\ell^A,A,\sigma)\to (\mathrm{e}^{\frac{\mathrm{1}}{2}(\lambda+i\phi)}\ell^A,A+ d(\lambda+i\phi),e^{\lambda+2i\phi}\sigma) $,
where $\phi$ is a U$(1)$ gauge parameter intrinsic to $\mathcal{N}$ and $\lambda$ generates internal boosts. 

At the null boundary $\mathcal{N}$, we now have to choose specific boundary conditions. At the kinematical level, the bulk fields $e_{AA'}$ and $\tnsr{A}{^A_B}$ are completely arbitrary. What is kept fixed is a gauge equivalence class of boundary fields,
\begin{equation}
g= [m,A,\theta]/_\sim,\qquad\delta g=0.
\end{equation}
The equivalence relations that define this equivalence class are diffeomorphisms of the null boundary, complexified U$(1)$ transformations, in addition to the following shift symmetries and conformal rescalings of the boundary data:
\begin{subequations}\label{null BCs}
\be
[m,A]\sim\big[m,A+\frac{\mathrm{i}}{2}\frac{\alpha}{\gamma+\mathrm{i}}\big],\label{shift1}\sim[m,A+\zeta\bar{m}+\bar{\zeta}m],
\sim[\mathrm{e}^f m,A],
\ee
\end{subequations}
where the 1-form $\alpha$ and the function $f$ are real, but $\zeta$ is complex. It is easy to check that this equivalence class is determined by two local degrees of freedom along $\mathcal{N}$, i.e.\ the two local degrees of freedom of gravitational radiation along a null boundary.

Given the boundary conditions \eref{null BCs}, the corresponding boundary action \cite{Wieland:2021vef} is
\begin{equation}
S_{\mathcal{N}}=\left[\frac{\mathrm{i}}{8\pi\gamma G}(\gamma+\mathrm{i})\int_{\mathcal{N}}\Big(e_A\wedge\bar{m}\wedge \Big(D-\frac{1}{2}A\Big)\ell^A-\frac{\theta}{4}e_A\wedge e^A\wedge\bar{m}\Big)\right]+\mathrm{cc}.
\end{equation}
This action has simple structure. The first term is just a gauged kinetic term for a configuration variable $\ell^A$, which is charged under SL$(2,\mathbb{C})\times \mathrm{U}_{\mathbb{C}}(1)$, and $e_A\wedge\bar{m}$ is the corresponding chiral momentum variable. The second term is a simple quadratic Hamiltonian. 
The coupled bulk plus boundary action is the sum of the two terms
\begin{equation}
S_{\text{bulk+boundary}} = S_{\mathcal{M}} + S_{\mathcal{N}}.\label{bulkbndryactn}
\end{equation}

Notice that the shift symmetry \eref{shift1} depends on the Barbero--Immirzi parameter $\gamma$. In fact, variations of the boundary action with respect to the boundary 1-form $\alpha\in\Omega^1(\mathcal{N}:\mathbb{R})$ impose a constraint, namely the \emph{reality condition}
\begin{equation}
\varepsilon = \bar{\varepsilon}.
\end{equation}

From the variation of the boundary spinors $e_A$ and $\ell^A$, we obtain additional boundary field equations that propagate the boundary spinors along $\mathcal{N}$. The resulting boundary field equations can be written in the following compact form
\begin{align}
\bar{m}\wedge \Big(D-\frac{1}{2}A\Big)\ell^A & = \frac{\theta}{2}e^A\wedge\bar{m},\label{bndryFeq1}\\
\Big(D+\frac{1}{2}A\Big)\wedge(e_A\wedge\bar{m}) & =0.\label{bndryFeq2}
\end{align}
If we take into account the decomposition of the spin connection with respect to the spin frame $(k^A,\ell^A)$, it is easy to check that these boundary field equations are satisfied on any embedded null surface. This is important to note, because it means that the boundary field equations impose no constraints on the free data along $\mathcal{N}$. In other words, the boundary field equations \eref{bndryFeq1} and \eref{bndryFeq2} can be satisfied on any 3-dimensional null boundary.




Given the boundary conditions and boundary field equations, we infer the symplectic potential. On a partial Cauchy surface $M$ that intersects $\mathcal{N}$ in a cross section $S$, there is a contribution from the bulk and an additional boundary contribution
\begin{equation}
\Theta_\Sigma = \frac{\mathrm{i}}{8\pi\gamma G}(\gamma+\mathrm{i})\left[\int_\Sigma\Sigma_{AB}\wedge\delta A^{AB}-\oint_{\partial M}e_A\wedge\bar{m}\,\delta\ell^A\right]+\mathrm{cc}.\label{thetaM}
\end{equation}

The expression for the symplectic structure on the null surface simplifies if we restrict ourselves to only those variations $\delta[\cdot]$ that preserve the ruling of the null surface, i.e.\ $\delta \ell^a\sim\ell^a$, see \cite{Wieland:2021vef}. Under this condition, we obtain
\begin{equation}
\Theta_{\mathcal{N}}=-\frac{1}{8\pi\gamma G}\int_{\mathcal{N}}\varepsilon\wedge\delta A^{(\gamma)}-\frac{\mathrm{i}}{8\pi\gamma G}\int_{\mathcal{N}}\Big[(\gamma+\mathrm{i})\ell_A D\ell^A\wedge\delta(k\wedge\bar{m})-\mathrm{cc}.\Big],\label{ThetaN}
\end{equation}
where $k_a\in\Omega^1(\mathcal{N}:\mathbb{R})$ is the Ehresmann connection such that $k_a\ell^a=1$. Notice that the expression for the symplectic potential has the familiar structure from the loop gravity holonomy flux algebra, with fluxes $\varepsilon$, $k\wedge m$, and $k\wedge\bar{m}$ that are dual to 1-forms or connections. In fact, the area 2-form $\varepsilon$ is dual to the U$(1)$ connection
\begin{equation}
A^{(\gamma)}=\Gamma+\gamma \omega.\label{U1ABconnctn}
\end{equation}
The connection $A^{(\gamma)}$ is  an abelian version of the SU$(2)$ Ashtekar--Barbero connection, namely the sum of the intrinsic U$(1)$ connection $\Gamma$ and the extrinsic curvature $\omega$. In fact, given the U$(1)$ gauge transformations, it is easy to check that $A^{(\gamma)}$ transforms as an abelian U$(1)$ connection. If, on the other hand, we rescale $A^{(\gamma)}$ by $\gamma^{-1}$, we obtain an abelian connection for the internal boost symmetry  $\ell^a\rightarrow\mathrm{e}^\lambda\ell^a$. 

\noindent Let us now discuss two important limiting cases.

 In the case of an isolated horizon $\mathcal{N}\simeq\mathbb{S}^2\times\mathbb{R}$, we have no gravitational radiation crossing the null boundary. Shear and expansion of the null generators $\ell^a$ vanish. Therefore, the second term in the expression for the pre-symplectic potential \eref{ThetaN} disappears, and we are left with the first term.  Since the 
area element $\varepsilon_{ab}$ is conserved, the entire geometry becomes effectively 2-dimensional.   
Given an arbitrary 2-dimensional cross-section $S\simeq \mathbb{S}^2$ of $\mathcal{N}$, the symplectic potential is now simply given by
\begin{equation}
\Theta_{\,\mathrm{IH}}=\frac{1}{8\pi\gamma G}\oint_{S}\varepsilon\, h^{-1}\delta h,
\end{equation}
where $h(x)= \mathrm{Pexp}(-i\int_{\gamma(\vec{x})} A^{(\gamma)}$ is the U$(1)$ holonomy along the null generator based at $\vec{x}$. 
This defines the phase space $[T^\ast \mathrm{SU}(1)]^{S}$. Notice that the area 2-form is the generator of U$(1)$ gauge transformations, hence area is quantized.

Another important limiting case is when we send $\mathcal{N}$ to null infinity $\mathcal{I}^\pm$. In this case, the symplectic potential \eref{ThetaN} returns the usual radiative phase space \cite{Wieland:2020gno}. 

\subsection{Spinor representation}

\noindent 
From the previous analysis we see that at the corner of null surface, where the $\Sigma$ intersects the null surface, we obtain the Poisson brackets
\begin{align}
\big\{\tilde{\pi}_A(\vec{x}),\ell^B(\vec{y})\big\}_{S}=\delta^B_A\tilde{\delta}_{S}(\vec{x},\vec{y}),\qquad
\big\{\tilde{\bar{\pi}}_{A'}(\vec{x}),\bar{\ell}^{B'}(\vec{y})\big\}_{S}=\delta^{B'}_{A'}\tilde{\delta}_{S}(\vec{x},\vec{y}),
\end{align}
where we introduced the momentum spinor $\tilde{\pi}_{A}$ as the pull-back of the boundary field $e_A\wedge \bar{m}$ to the cross-section $S$, i.e.
\begin{equation}
\tilde{\pi}_A = \frac{\mathrm{i}}{8\pi\gamma G}(\gamma+\mathrm{i})\varphi^\ast_{S}(e_A\wedge\bar{m}).
\end{equation}
A number of Dirac observables can be written in terms of these boundary fields. For example, the generator of infintesimal tangential diffeomorphism associated with  $\xi^a\in T\Sigma$, is given  by
\begin{equation}
P_\xi = \oint_{S}\xi^a[\tilde{\pi}_AD_a\ell^A]+\mathrm{cc}.
\end{equation}
In the same way, we find the generators of complexified U$(1)$ transformations,
\begin{align}
\tilde{\pi}_A\ell^A=\tilde{K}+\mathrm{i}\,\tilde{L}& = \frac{\gamma+\mathrm{i}}{8\pi \gamma G}\varepsilon,
\end{align}
where $\tilde{L}$ is a $U(1)$ generator, and $\tilde{K}$ is a boost generator and $\varepsilon$ is the area form. In terms of harmonic oscillators, $\tilde{L}$ is the difference of two number operators, and $\tilde{K}$ is a two-mode squeeze operator, see \cite{Wieland:2017cmf}. The spectrum of $\tilde{K}$ is continuous, the spectrum of $\tilde{L}$ is discrete. 
The area 2-form \eref{epsdef} is the contraction of the boundary spinors, 
%
At the kinematical level, this surface density is complex. For the area to be real,  we have to impose the simplicity constraint
$
\tilde{K}-\gamma \tilde{L} = 0$.

When acting on any physical state, the area density will be proportional to $\tilde{L}$, which has a discrete spectrum. Hence area is quantized.  This selects a representation of the canonical commutation relations. In this way, the quantization of area on the null surface boundary agrees with our analysis in section \ref{sec:representation}. Both results only rely on the compactness of the orbits generated by the area density $\varepsilon\propto\tilde{L}$. If we use a spin network representation, the boundary spinors $(\tilde{\pi}_A,\ell^A)$ will be excited at a number of punctures. We could also use, however, a more standard Fock representation \cite{Wieland:2017cmf}, where $\tilde{\pi}_A$ and $\ell^A$ are continuous. In either case, the result is the same. To summarise the the Barbero\,--\,Immirzi parameter, deforms the boundary symmetries in such a way that the area spectrum becomes discrete. 
\section{Summary and perspectives}

In this review, we have discussed three main ideas and related results. The first result is about the classical phase space of general relativity in the presence of inner boundaries of a spacelike or null slice. When there is a boundary, an otherwise unphysical gauge redundancy in the bulk turns into a physical degree of freedom intrinsic to the boundary. The corresponding conjugate elements on phase space are the canonical generators of the boundary symmetries. It is for this reason that the generators of diffeomorphisms turn into surface charges that measure quasi-local observables of the gravitational field. These observables generalise the ten Minkowskian Noether charges for the Poincaré group into an infinite set of boundary charges that satisfy the so-called corner symmetry algebra. We then reviewed a number of past results from loop quantum gravity, where the boundary conditions and resulting boundary symmetries have led to important results, from local entanglement \cite{BarberoG:2022ixy} to the entropy counting of black hole horizons \cite{Bianchi:2023avf}, and to renormalization and the continuum limit of the theory \cite{Asante:2022dnj}. Then, we explained one of the key results of the corner symmetry program thus far, namely the fact that the loop quantization of area can be understood from the quantization of the boundary modes alone \cite{Wieland:2021vef,Wieland:2017cmf,Freidel:2020ayo}. Since there is a unique area element on a null surface, the construction becomes particularly transparent for such boundaries. The area 2-form at a 2-dimensional cross section turns into a surface charge, whose spectrum is discrete. From the perspective of the gravitational phase space in a finite region, this surface charge is a complete Dirac observable. The discreteness of the spectrum of the corresponding quantum operator can be traced back to the presence of the Barbero--Immirzi parameter in the action. The Barbero--Immirzi parameter does not affect the commutation relations for the radiative modes in the bulk, but it deforms the algebra of boundary symmetries and their generators. This discreteness has been proven to  follows from representing the corner symmetry as in \eqref{rhorep}. Finally, we have also reviewed a number of results on how the Einstein equations induce a field theory for a Carrollian fluid on the null surface boundary and how this can also be described naturally within the tetrad formulation used in loop quantum gravity. This furthermore connects seamlessly with the description of isolated horizons developed in the LQG black hole entropy counting.

Some developments were not covered in our review. For example, there is a growing body of research in quantum foundations and quantum information science on quantum reference frames and their relation with edge modes \cite{Goeller:2022rsx,Vanrietvelde:2018pgb,Carrozza:2022xut,Kabel:2023jve}. Quantum reference frames are dual (in the phase space sense of the word) to the generators of coordinate transformations. Recent results by Brukner and collaborators gave an operational definition of \emph{quantum reference frames}, see e.g. \cite{Giacomini:2017zju}. This definition includes a prescription for how to \emph{jump} between quantum reference frames and identify properties of a quantum state that remain invariant under such transformations. It turns out, for example, that quantum entanglement is an observer-dependent notion and depends on the frame of reference. Stenghthening further the connection between corner symmetry and quantum reference frames is a promising area of development. 


Altogether, the idea we arrive at is that of local holography. While AdS holography on asymptotic timelike boundaries is made possible by reflecting boundary conditions, which prevent from including radiation \cite{Compere:2020lrt}. Celestial-carrollian Holography allows the inclusion of radiation, and is adapted to the description of physically realistic setups \cite{Strominger:2017zoo,Raclariu:2021zjz,Pasterski:2021rjz, Fareghbal:2013ifa,Donnay:2022aba,Donnay:2021wrk}. Instead of considering null or timelike boundaries at infinity, local holography aims at obtaining constraints on quantum gravity from the local corner symmetry structure associated with any arbitrary subregion defined by an entangling cut. This idea should not necessarily be viewed as a new proposal for quantum gravity, but rather as a consistency framework that brings a universal bottom-up perspective for all candidate approaches. In this review, we have discussed the lessons from and for loop quantum gravity, but it will be interesting and important to discuss local holography in other contexts as well.

\bibliography{library}

\end{document}